\documentclass[prd,twocolumn]{revtex4}
\usepackage{graphicx, epsfig}
\usepackage{color}
\usepackage{mathrsfs}

\newcommand{\be}{\begin{equation}}
\newcommand{\ee}{\end{equation}}
\newcommand{\bea}{\begin{eqnarray}}
\newcommand{\eea}{\end{eqnarray}}

\newcommand{\gapp}{\mathrel{\raise.3ex\hbox{$>$}\mkern-14mu
              \lower0.6ex\hbox{$\sim$}}}
\newcommand{\lapp}{\mathrel{\raise.3ex\hbox{$<$}\mkern-14mu
              \lower0.6ex\hbox{$\sim$}}}
\newcommand{\scri}{\mathscr{I}}

\begin{document}
\title{Observation of Incipient Black Holes and the Information Loss Problem}
\author{Tanmay Vachaspati}
\author{Dejan Stojkovic}
\affiliation{CERCA, Department of Physics, 
Case Western Reserve University, Cleveland, OH~~44106-7079}
\author{Lawrence M. Krauss}
\affiliation{CERCA, Department of Physics, 
Case Western Reserve University, Cleveland, OH~~44106-7079 (permanent address)}
\affiliation{also, Department of Physics and Astronomy, Vanderbilt University, Nashville, TN}

\begin{abstract}
\noindent
We study the formation of black holes by spherical domain wall 
collapse as seen by an asymptotic observer, using the functional
Schrodinger formalism.
To explore what signals such observers will see, we study 
radiation of a scalar quantum field in the collapsing
domain wall background. 
The total energy flux radiated diverges when backreaction of the 
radiation on the collapsing wall is ignored, and the domain wall 
is seen by the asymptotic observer to evaporate by non-thermal 
``pre-Hawking radiation'' during the collapse process. 
Evaporation by pre-Hawking radiation implies that 
an asymptotic observer can never lose objects down a black hole. 
Together with the non-thermal nature of the radiation,
this may resolve the black hole information loss problem. 
\end{abstract}

\maketitle

\section{Introduction}
\label{introduction}

Black holes embody the long-standing theoretical challenge of 
combining general relativity and quantum mechanics, with various
proposals being advocated over the years to resolve paradoxes 
associated with black hole formation, evaporation and information loss. 
Resolution of these issues has become even more timely with the 
possible formation and evaporation of black holes in particle 
accelerators in the framework of higher dimensional models that 
have recently garnered much attention. 
The process of black hole formation is generally
discussed from the viewpoint of an infalling observer. However, 
in all physical settings it is the viewpoint of the asymptotic 
observer that is relevant. More concretely, if a black hole is 
formed in the Large Hadron Collider, it has to be observed by 
physicists sitting on the CERN campus. The physicists have clocks 
in their offices and they watch the process of formation and
evaporation in this coordinate frame. They must address 
questions such as: At what time did a black hole form? 
Is any information lost into the black hole? How long did it take 
for the black hole to evaporate? What is the spectrum of the decay 
products? 

The process of gravitational collapse has been studied extensively
over the last few decades, from many different viewpoints, including
1+1 dimensional models and modifications of general relativity
({\it e.g.} see \cite{BirrellandDavies}).
Unlike a large subset of this work, our analysis is in 3+1 
dimensions and within conventional general relativity. We model 
the general problem by choosing to study a collapsing 
spherical shell of matter, more specifically a vacuum domain wall. 
The physical setup of the problem and the functional Schrodinger
formalism are described in Sec.~\ref{setup}.

A crucial aspect of our analysis is that we address the question 
of black hole formation and evaporation {\em  as seen by an asymptotic 
observer}. Initially, when the domain wall is large, the spacetime
is described by the Schwarzschild metric, just as for a static
star. From here on, the wall and the metric are evolved forward
in time, always using the Schwarzschild time coordinate.
We emphasize that all our discussion, unless explicitly stated,
refers to the Schwarzschild time, $t$, and this defines the time 
slicing of the spacetime. As is well known, the Schwarzschild coordinate 
system breaks down at a black hole horizon, and there is danger that 
our analysis will also break down at some point during the 
gravitational collapse. However, we do not encounter any such 
difficulties, suggesting that our calculation is self-consistent. 
A second danger is that the coordinate system may provide an 
incomplete description of the gravitational collapse spacetime. 
This remains a possibility. However, we find that Schwarzschild 
coordinates are sufficient to answer the very specific set of questions 
we ask from the asymptotic observer's viewpoint. Namely, does the 
asymptotic observer see objects disappear into a black hole in the 
time that he sees the collapsing body evaporate? And, is the spectrum 
of the radiation received ever truly thermal (even in the semiclassical
approximation)?

In Sec.~\ref{classical} we verify the standard result that 
the formation of an event horizon takes an infinite (Schwarzschild)
time if we consider classical collapse. 
This is not surprising and is often viewed as a 
limitation of the Schwarzschild coordinate system. 
To see if this result changes 
when quantum effects are taken into account, we address the problem 
of quantum collapse using a minisuperspace version of the functional 
Schrodinger equation \cite{DeWitt} in Sec.~\ref{qcollapse}. We find 
that even in this case the black hole takes an infinite time to form,
contrary to some speculations in the literature \cite{frolov}. 

In Sec.~\ref{semiclassradn} we consider the possible radiation
associated with the collapsing shell by considering the 
interaction of a quantum scalar field and the classical background
of a collapsing domain wall. We treat the problem using the
functional Schrodinger picture, which we relate to the standard
Bogolubov treatment carried out in Sec.~\ref{appa}. 
Here we find that the shell, even as it collapses,
radiates away its energy in a finite amount of time. With some 
assumptions about
the metric close to the incipient horizon, we conclude that the
evaporation time is shorter than what would be taken by objects 
to fall through a black hole horizon. This leads us to the
conclusion that the asymptotic observer will see the evaporation
of the collapsing shell before he can see any objects disappear.

We discuss our results from the point of view of an infalling 
observer in Sec.~\ref{infallingobserver}, where we attempt to reconcile 
the fact that such an observer will not see substantial radiation
with the observations made by an asymptotic observer.
Our conclusions are summarized in Sec.~\ref{discussion}, where we 
elucidate a possible resolution of the information loss problem 
suggested by our results, together with a discussion of possible 
loopholes and future directions.

\section{Setup and formalism}
\label{setup}

To study a concrete realization of black hole formation we consider a 
spherical Nambu-Goto domain wall that is collapsing. To include 
the possibility of (spherically symmetric) radiation we consider a
massless scalar field, $\Phi$, that is coupled to the gravitational
field but not directly to the domain wall. The action for
the system is
\begin{eqnarray}
S &=& \int d^4 x \sqrt{-g} \left [ -\frac{1}{16\pi G} {\cal R}
     + \frac{1}{2} (\partial _\mu \Phi )^2 \right ] \nonumber \\
     &-& {\sigma} \int d^3 \xi \sqrt{- \gamma} + S_{\rm obs}
\label{action}
\end{eqnarray}
where the first term is the Einstein-Hilbert action for the
gravitational field, the second is the scalar field action,
the third is the domain wall action in terms of the wall
world volume coordinates, $\xi^a$ ($a=0,1,2$), the wall
tension, ${\sigma}$, and the induced world volume metric
\begin{equation}
\gamma_{ab} = g_{\mu\nu} \partial_a X^\mu \partial_b X^\nu
\label{inducedmetric}
\end{equation}
The coordinates $X^\mu (\xi^a )$ describe the location of the wall
and Roman indices go over internal domain wall world-volume coordinates
$\zeta^a$, while Greek indices go over space-time coordinates. 
The term $S_{\rm obs}$ in Eq.~(\ref{action}) denotes the action 
for the observer.

We will begin first with the Wheeler-de Witt equation in order 
to explore and contrast quantum vs classical collapse of the domain wall, 
but we will
eventually utilize the functional Schrodinger formalism to study
both collapse and radiation in this system.  
 
The Wheeler-de Witt equation for a closed universe is
\begin{equation}
H \Psi = 0
\end{equation}
where $H$ is the Hamiltonian and $\Psi [X^\alpha,g_{\mu\nu},\Phi,{\cal O}]$ 
is the wave-functional for all the ingredients of the system, including the
observer's degrees of freedom denoted by ${\cal O}$. Note that the 
wave-functional, $\Psi$, is a functional of the fields but not of the 
spacetime coordinates. 
We will separate the Hamiltonian into two parts, one for the system 
and the other for the observer
\begin{equation}
H = H_{\rm sys} + H_{\rm obs}
\end{equation}
Any (weak) interaction terms between the observer and the wall-metric-scalar 
system are included in $H_{\rm sys}$. 
The observer is assumed not to significantly affect the evolution of the 
system and similarly for the system vis a vis the observer. The total wave-functional 
can be written as a sum over eigenstates
\begin{equation}
\Psi = \sum_k c _k \Psi^k_{\rm sys} ({\rm sys},t) \Psi^k_{\rm obs} ({\cal O},t)
\end{equation}
where $k$ labels the eigenstates, $c_k$ are complex coefficients, and we 
have introduced the observer time, $t$, via
\begin{equation}
i \frac{\partial \Psi^k_{\rm obs}}{\partial t} 
           \equiv H_{\rm obs} \Psi^k_{\rm obs}
\end{equation}
With the help of an integration by parts, and the fact that the total
wave-functional is independent of $t$, the Wheeler-de Witt equation implies 
the Schrodinger equation
\begin{equation}
H_{\rm sys} \Psi^k_{\rm sys} = 
i \frac{\partial \Psi^k_{\rm sys}}{\partial t} 
\label{schrodingerpre}
\end{equation} 
For convenience, from now on we will denote the system wave-function
simply by $\Psi$ and drop the superscript $k$ and the subscript 
``${\rm sys}$''. Similarly $H$ will now denote $H_{\rm sys}$, and the
Schrodinger equation reads 
\begin{equation}
H \Psi = i \frac{\partial \Psi}{\partial t} 
\label{schrodinger}
\end{equation} 

A general treatment of the full Wheeler-de Witt equation is very
difficult and we shall utilize the frequently employed strategy of
truncating the field degrees of freedom to a finite subset. In other
words, we will consider a minisuperspace version of the Wheeler-de Witt
equation. As long as we keep all the degrees of freedom that are of
interest to us, this is a useful truncation. With this in mind,
we only consider spherical domain walls and assume spherical symmetry
for all the fields. So the wall is described by only the radial 
degree of freedom, $R(t)$. The metric is taken to be the solution 
of Einstein equations for a spherical domain wall. 
The metric is Schwarzschild outside the wall, as follows 
from spherical symmetry \cite{Ipser:1983db}
\begin{equation}
ds^2= -(1-\frac{R_S}{r}) dt^2 + (1-\frac{R_S}{r})^{-1} dr^2 + 
      r^2 d\Omega^2 \ , \ \ r > R(t)  
\label{metricexterior}
\end{equation}
where, $R_S = 2GM$ is the Schwarzschild radius in terms of the mass,
$M$, of the wall, and
\begin{equation}
d\Omega^2  = d\theta^2  + r^2 \sin^2\theta d\phi^2 
\end{equation}
In the interior of the spherical domain wall, the line element
is flat, as expected by Birkhoff's theorem, 
\begin{equation}
ds^2= -dT^2 +  dr^2 + r^2 d\theta^2  + r^2 \sin^2\theta d\phi^2  \ , 
\ \ r < R(t) 
\label{metricinterior}
\end{equation}
The interior time coordinate, $T$, is related to the observer time 
coordinate, $t$, via the proper time, $\tau$, of the domain wall.
\begin{equation}
\frac{dT}{d\tau} = 
      \left [ 1 + \left (\frac{dR}{d\tau} \right )^2 \right ]^{1/2}
\label{bigTandtau}
\end{equation}
and
\begin{equation}
\frac{dt}{d\tau} = \frac{1}{B} \left [ B + 
         \left ( \frac{dR}{d\tau} \right )^2 \right ]^{1/2}
\label{littletandtau}
\end{equation}
where 
\begin{equation}
B \equiv 1 - \frac{R_S}{R}
\label{BofR}
\end{equation}
The ratio of these equations gives
\begin{equation}
\frac{dT}{dt} = \frac{(1+R_{\tau}^2)^{1/2} B}{(B + R_{\tau}^2)^{1/2}}
= \left [ B - \frac{(1-B)}{B} {\dot R}^2 \right ]^{1/2}
\label{tT}
\end{equation}
where $R_{\tau} = dR/d\tau$ and ${\dot R} = dR/dt$. 
Integrating Eq.~(\ref{tT}) still requires 
knowing $R(\tau)$ (or $R(t)$) as a function of $\tau$ (or $t$).

Since we are restricting our minisuperspace to fields with spherically 
symmetry, we need not include any other metric degrees of freedom.
The scalar field can also be truncated to the spherically symmetric
modes ($\Phi = \Phi (t,r)$). 

By integrating the equations of motion for the spherical domain wall,
Ipser and Sikivie \cite{Ipser:1983db} found that the mass is a constant 
of motion and is given by
\begin{equation}
M = \frac{1}{2} [ \sqrt{1+R_\tau^2} + \sqrt{B+ R_\tau^2} ] 4\pi \sigma R^2 
\label{ISmass}
\end{equation}
where it is assumed that ${\rm max}(R) > 1/4\pi G\sigma$ 
\cite{Ipser:1983db}.
This expression for $M$ is implicit since $R_S =2GM$ occurs in $B$.
Solving for $M$ explicitly in terms of $R_\tau$ gives
\begin{equation}
M = 4\pi \sigma R^2 [ \sqrt{1+R_\tau^2} - 2\pi G\sigma R]
\label{MRtau}
\end{equation}
and with the relations between $T$ and $\tau$ given above we get
\begin{equation}
M = 4\pi \sigma R^2 \left [ \frac{1}{\sqrt{1-R_T^2}} - 2\pi G\sigma R \right ]
\label{mconserve}
\end{equation}
where $R_T$ denotes $dR/dT$.

We now discuss the classical collapse of the domain wall.

\section{Classical treatment of domain wall collapse}
\label{classical}

A naive approach to obtaining the dynamics for the spherical domain 
wall is to insert the spherical ansatz for both the wall and the metric 
in the original action. This would lead to an effective action for the 
radial coordinate $R(t)$. However, it is known that this approach
does not give the correct dynamics for gravitating systems.
We find that this approach does not straightforwardly lead to 
mass conservation as given in Eq.~(\ref{ISmass}). So we take the 
alternative approach of finding an action that leads to
the correct mass conservation law. The form of the action can
be deduced from Eq.~(\ref{mconserve}) quite easily
\begin{equation}
S_{eff} = - 4\pi \sigma \int dT R^2 
        \left [ \sqrt{1-R_T^2} - 2\pi G\sigma R \right ]
\end{equation}
which can now be written in terms of the external time $t$
\begin{eqnarray}
S_{eff} &=& - 4\pi \sigma \int dt R^2 
        \biggl [ \sqrt{B-\frac{{\dot R}^2}{B}}  \nonumber \\
        &-& 2\pi G \sigma R 
                \sqrt{B - \frac{1-B}{B} {\dot R}^2} ~ \biggr ]
\label{Seff}
\end{eqnarray}
and the effective Lagrangian is
\begin{equation}
L_{eff} = - 4\pi \sigma R^2 
        \left [ \sqrt{B-\frac{{\dot R}^2}{B}} - 2\pi G \sigma R 
                \sqrt{B - \frac{1-B}{B} {\dot R}^2} \right ]
\label{efflagrangian}
\end{equation}

The generalized momentum, $\Pi$, can be derived from 
Eq.~(\ref{efflagrangian}) 
\begin{equation}
\Pi = \frac{4\pi \sigma R^2 {\dot R}}{\sqrt{B}} \left [
      \frac{1}{\sqrt{B^2-{\dot R}^2}} - 
       \frac{2\pi G\sigma R (1-B)}{\sqrt{B^2 - (1-B) {\dot R}^2}}
                        \right ] 
\label{momentum}
\end{equation}
The Hamiltonian (in terms of ${\dot R}$) is
\begin{equation}
H = 4\pi \sigma B^{3/2}R^2 \left [
         \frac{1}{\sqrt{B^2-{\dot R}^2}} - 
          \frac{2\pi G\sigma R}{\sqrt{B^2- (1-B){\dot R}^2}}
                                  \right ]
\label{Ham}
\end{equation}

To obtain $H$ as a function of $(R, \Pi )$, we need to eliminate
${\dot R}$ in favor of $\Pi$ using Eq.~(\ref{momentum}). This can
be done but is messy, requiring solutions of a quartic polynomial. 
Instead we consider the limit when $R$ is
close to $R_S$ and hence $B \rightarrow 0$. In this limit the
denominators of the two terms in Eqs.~(\ref{momentum}) (also in
(\ref{Ham})) are equal and
\begin{equation}
\Pi \approx \frac{4\pi \mu R^2  {\dot R}}
              {\sqrt{B} \sqrt{B^2-{\dot R}^2}}
\end{equation}
where
\begin{equation}
\mu \equiv \sigma (1-2\pi G\sigma R_S)
\end{equation}
and
\begin{eqnarray}
H &\approx& \frac{4\pi \mu B^{3/2}R^2}{\sqrt{B^2-{\dot R}^2}} \label{HRdot}\\
  &=& \left [  (B\Pi)^2 + B (4\pi \mu R^2)^2 \right ] ^{1/2} \label{HPi}
\end{eqnarray}
The Hamiltonian has the form of the energy of a relativistic
particle, $\sqrt{p^2 + m^2}$, with a position dependent mass. 

The Hamiltonian is a conserved quantity and so, from Eq.~(\ref{HRdot}),
\begin{equation}\label{defh}
\frac{ B^{3/2}R^2}{\sqrt{B^2-{\dot R}^2}} =h
\end{equation}
where $h = H/4\pi \mu $ is a constant. (Up to the approximation
used to obtain the simpler form of the Hamiltonian in Eq.~(\ref{HRdot}),
the Hamiltonian is the mass.)

Solving Eq. (\ref{defh}) for ${\dot R}$ we get
\begin{equation}
{\dot R} = \pm B \left(1- \frac{BR^4}{h^2} \right)^{1/2}\, ,
\label{Rdotsolution}
\end{equation}
which, near the horizon, takes the form 
\begin{equation}
{\dot R} \approx \pm B\left(1- {1 \over 2} \frac{BR^4}{h^2} \right)
\label{rdotnh}
\end{equation}
since $B \rightarrow 0$ as $R \rightarrow R_S$.

The dynamics for $R \sim R_S$ can be obtained by solving the equation 
${\dot R} =  \pm B$. To leading order in $R-R_S$, the solution is
\begin{equation}
R(t) \approx R_S + (R_0-R_S) e^{\pm t/R_S} \, .
\label{solution}
\end{equation}
where $R_0$ is the radius of the shell at $t=0$. As we are interested 
in the collapsing solution, we choose the negative sign in the exponent. 
This solution implies that, from the classical point 
of view, the asymptotic observer never sees the formation of the horizon
of the black hole, since $R(t) = R_S$ only as $t \rightarrow \infty$.
This result is similar to the well-known result (for example, 
see \cite{Townsend}) that it takes an infinite time for objects to 
fall into a pre-existing black hole as viewed by an asymptotic 
observer \cite{MTW}. 
In our case there is no pre-existing horizon, which is itself taking 
an infinite amount of time to form during collapse.
To see if 
this conclusion will change when quantum effects are taken into 
account ({\it e.g.} Sec.~10.1.5 of \cite{frolov}) we now explore
the quantum dynamics of the collapsing domain wall.

\section{Quantum treatment of domain wall collapse}
\label{qcollapse}

The classical Hamiltonian in Eq.~(\ref{HPi}) has a square root and
so we first consider the squared Hamiltonian
\begin{equation}\label{Hsq}
H^2 = B\Pi ~B \Pi + B (4\pi \mu R^2)^2 
\end{equation}
where we have made a choice for ordering $B$ and $\Pi$ in
the first term. In general, we should add terms that depend on
the commutator $[B, \Pi ]$. However, in the limit $R\rightarrow
R_S$, we find
$$
[B, \Pi] \sim \frac{1}{R_S}
$$
Estimating $H$ by the mass, $M$, of the domain wall, the
terms due to the operating order ambiguity will be negligible 
provided
$$
M \gg \frac{1}{R_S} \sim \frac{m_P^2}{M}
$$
where $m_P$ is the Planck mass. Hence the operator ordering
ambiguity can be ignored for domain walls that are much more 
massive than the Planck mass.

Now we apply the standard quantization procedure. We substitute 
\begin{equation}
\Pi = -i \frac{\partial}{\partial R}
\end{equation}
in the squared Schrodinger equation, 
\begin{equation}
H^2 \Psi = - \frac{\partial^2 \Psi}{\partial t^2}
\label{Hsquared}
\end{equation}
Then
\begin{equation}\label{Hsqexplicit}
- B\frac{\partial}{\partial R}
 \left ( B \frac{\partial \Psi}{\partial R} \right ) + 
    B (4\pi \mu R^2)^2 \Psi = - \frac{\partial^2 \Psi}{\partial t^2}
\end{equation}
To solve this equation, define
\begin{equation}
u = R + R_S \ln \left | \frac{R}{R_S} - 1 \right |
\label{uandR}
\end{equation}
which gives
\begin{equation}
B\Pi = -i \frac{\partial}{\partial u}
\end{equation}
Eq.~(\ref{Hsquared}) then gives 
\begin{equation}
  \frac{\partial^2 \Psi}{\partial t^2}
 - \frac{\partial^2 \Psi}{\partial u^2} 
 + B (4\pi \mu R^2)^2 \Psi = 0
\label{waveeq}
\end{equation}
This is just the massive wave equation in a Minkowski background
with a mass that depends on the position. Note that $R$ needs to
be written in terms of the coordinate $u$ and this can be done
(in principle)
by inverting Eq.~(\ref{uandR}). However, care needs to be taken
to choose the correct branch since the region 
$R \in (R_S,\infty)$ maps onto $u\in (-\infty ,+\infty)$
and $R \in (0,R_S)$ onto $u \in (0,-\infty)$.

We are interested in the situation of a collapsing wall. 
In the region $R \sim R_S$, the logarithm in Eq.~(\ref{uandR})
dominates and 
$$
R \sim R_S + R_S e^{u/R_S}
$$
We look for wave-packet solutions propagating toward 
$R_S$, that is, toward $u \rightarrow - \infty$. In this limit
$$
B \sim e^{u/R_S} \rightarrow 0
$$
and the last term in Eq.~(\ref{waveeq}) can be ignored.
Wave packet dynamics in this region is simply given by the
free wave equation and any function of light-cone coordinates
($u\pm t$) is a solution. In particular, we can write a Gaussian
wave packet solution that is propagating toward the Schwarzschild
radius 
\begin{equation}
\Psi = \frac{1}{\sqrt{2\pi} s} e^{- (u+t)^2/2s^2 }
\label{packetsolution}
\end{equation}
where $s$ is some chosen width of the wave packet in the $u$
coordinate. The width of the Gaussian wave packet remains fixed in 
the $u$ coordinate while it shrinks in the $R$ coordinate via the 
relation $dR = B du$ which follows from Eq.~(\ref{uandR}).   This fact
is of great importance, since if the wave packet remained of constant size
in $R$ coordinates, it might cross the event horizon in finite time.

The wave packet travels at the speed of light in the $u$ coordinate 
-- as can be seen directly from the wave equation Eq.~(\ref{waveeq}) 
or from the solution, Eq.~(\ref{packetsolution}). However, to get to the
horizon, it must travel out to $u=-\infty$, and this takes an infinite
time. So we conclude that the quantum domain wall does not collapse 
to $R_S$ in a finite time, as far as the asymptotic observer is concerned,
so that quantum effects do not alter the classical result that
an asymptotic observer does not observe the formation of an event 
horizon.

The above analysis shows that the collapsing wall takes an infinite
time to reach $R=R_S$. The analysis leaves room for processes by which 
the wave packet can jump from the $(R_S,\infty)$ region to the $(0,R_S)$ 
region, without ever going through $R_S$. Note that this is different
from tunneling through a barrier. In that case, the wave function is
non-zero within the barrier, and a small part of it leaks over to
the other side of the barrier. In the present case, $R_S$ occurs
at $u=-\infty$ and so, if there is any barrier, it is infinitely
far away. If there is to be a jump from outside to inside $R_S$,
it does not show up in the present description using the Wheeler-de Witt 
equation.

We have obtained the massive wave equation, Eq.~(\ref{waveeq}), by
first squaring the classical Hamiltonian, Eq.~(\ref{HPi}). This
procedure eliminated the square root occurring in the Hamiltonian.
It is possible that some other quantization procedure will yield
different conclusions. In this context, we note, in fact, that we need
not square the Hamiltonian to get rid of the square root provided
we work in the near horizon limit. In that case
\begin{equation}
H = \left [  (B\Pi)^2 + B (4\pi \mu R^2)^2 \right ] ^{1/2} 
 \approx  \pm B\Pi 
\label{HBPi}
\end{equation}
where the sign is chosen to make $H$ non-negative.
Then the Schrodinger equation again yields wave packets propagating
at the speed of light in the $(t,u)$ coordinate system and with the
horizon located at $u=-\infty$.

\section{Radiation - semiclassical treatment}
\label{semiclassradn}

If an external observer never sees the formation of an event horizon, 
we need to explore what radiation might be observed that characterizes 
gravitational collapse. To do so we consider a quantum scalar field 
in the background of the collapsing domain wall. We do not consider
gravitational radiation since this is excluded by our restriction 
to spherically symmetric modes in minisuperspace. 
In this section, 
we approach the problem using the functional Schrodinger equation
since (i) we have already set up this approach and used it in the 
previous section, (ii) we believe the approach is more suited
to treating the backreaction problem, and (iii) it allows us
to calculate the total radiation of which Hawking
radiation may only be a subset. To connect with earlier
work, we discuss the problem of Hawking radiation using the 
conventional Bogolubov transformations in Sec.~\ref{appa}. 

The action for the scalar field is
\begin{equation}
S = \int d^4x \sqrt{-g} \frac{1}{2} g^{\mu \nu}
                 \partial_\mu \Phi \partial_\nu \Phi
\end{equation}

We decompose the (spherically symmetric) scalar field into a complete set
of real basis functions denoted by $\{ f_k (r) \}$
\begin{equation}
\Phi = \sum_k  a_k(t) f_k (r) 
\label{modes}
\end{equation}
The exact form of the functions $f_k (r)$ will not be important for us. 
We will be interested in the wavefunction for the mode coefficients
$\{ a_k \}$. 

In the functional Schrodinger picture, we wish to find the
wavefunctional $\Psi [ \Phi ; t ]$ by solving a functional
Schrodinger equation. This is equivalent to finding the
wavefunction of an infinite set of variables, 
$\psi ( \{ a_k \} , t )$, by solving an ordinary Schrodinger
equation in an infinite dimensional space. The 
procedure (detailed below) is to find independent eigenmodes, 
$\{ b_k \}$, for the system for which the Hamiltonian is a sum 
of terms, one for each independent eigenmode. Then the wavefunction
factorizes and can be found by solving a time-dependent Schrodinger 
equation of just one variable.

Since the metric inside and outside the shell have different forms, 
we split the action into two parts
\begin{equation}
S = S_{\rm in} + S_{\rm out}
\end{equation}
where
\begin{equation}
S_{\rm in} = 2\pi \int dt \int_0^{R(t)} \hskip -0.1 in dr ~ r^2 
 \left [
 - \frac{(\partial_t \Phi)^2}{\dot T} + {\dot T} (\partial_r \Phi)^2 
 \right ] 
\label{actionterms1} 
\end{equation}
\begin{eqnarray}
S_{\rm out} &=& 2\pi \int dt \int_{R(t)}^\infty dr ~ r^2 
 \biggl [
 - \frac{(\partial_t \Phi)^2}{1-R_S/r} \nonumber \\
 && \hskip 0.75in +\left ( 1-\frac{R_S}{r} \right ) (\partial_r \Phi)^2 
  \biggr ]
\label{actionterms2}
\end{eqnarray}
${\dot T}$ is given by Eq.~(\ref{tT}), which with 
Eq.~(\ref{Rdotsolution}), gives 
\begin{equation}
{\dot T} = \frac{dT}{dt} 
         = B \left [ 1 + (1-B)\frac{R^4}{h^2} \right ]^{1/2}
\end{equation}
As $R \rightarrow R_S$, ${\dot T} \sim B \rightarrow 0$. Therefore
the kinetic term in $S_{\rm in}$ diverges as $ (R-R_S)^{-1}$ in this 
limit and dominates over the softer logarithmically divergent 
contribution to the kinetic term from $S_{\rm out}$. Similarly
the gradient term in $S_{\rm in}$ vanishes in this limit and is
sub-dominant compared to the contribution coming from $S_{\rm out}$. Hence,
\begin{eqnarray}
S &\sim& 2\pi \int dt \biggl [
 - \frac{1}{B} \int_0^{R_S} dr ~r^2 (\partial_t \Phi)^2 \nonumber \\
 && \hskip 0.25in + \int_{R_S}^\infty dr ~r^2 \left ( 1-\frac{R_S}{r} \right )
                              (\partial_r \Phi )^2 
                    \biggr ]
\label{effaction}
\end{eqnarray}
where we have changed the limits of the integrations to $R_S$
since we are working in the regime $R(t) \sim R_S$. This approximation
is valid provided the contribution from $r \in (R_S, R(t))$ to the
integrals 
remains subdominant, and also the time variation introduced by
the true integration limit ($R(t)$) can be ignored. These requirements 
are not arduous. 

Now, we use the expansion in modes in Eq.~(\ref{modes}) to write
\begin{equation}
S = \int dt \left [ - \frac{1}{2B} {\dot a}_k {\bf M}_{kk'} {\dot a}_{k'} 
                     + \frac{1}{2} a_k {\bf N}_{kk'} a_{k'} \right ]
\end{equation}
where ${\bf M}$ and ${\bf N}$ are matrices that are independent of 
$R(t)$ and are given by
\begin{equation}
{\bf M}_{kk'} = 4\pi \int_0^{R_S} dr ~ r^2 f_k (r) f_{k'} (r)
\label{M11}
\end{equation}
\begin{equation}
{\bf N}_{kk'} = 4\pi \int_{R_S}^\infty dr ~  
                     r^2 \left ( 1 - \frac{R_S}{r} \right )
                                   f_k ' (r) f_{k'} ' (r)
\label{N11}
\end{equation}

Using the standard quantization procedure, the wave function $\psi (a_k,t)$
satisfies
\begin{equation}
\biggl [ 
\left ( 1-\frac{R_S}{R} \right ) 
                \frac{1}{2} \Pi_k ({\bf M}^{-1})_{kk'} \Pi_{k'} 
 + \frac{1}{2} a_k {\bf N}_{kk'} a_{k'} \biggr ] \psi = 
i \frac{\partial \psi}{\partial t}
\end{equation}
where
\begin{equation}
\Pi_k = -i \frac{\partial}{\partial a_k}
\end{equation}
is the momentum operator conjugate to $a_k$.

So the problem of radiation from the collapsing domain wall is equivalent
to the problem of an infinite set of coupled harmonic oscillators whose
masses go to infinity with time. Since the matrices ${\bf M}$ and
${\bf N}$ are symmetric and real ({\it i.e.} Hermitian), it is
possible to do a principal axis transformation to simultaneously 
diagonalize ${\bf M}$ and ${\bf N}$ (see Sec.~6-2 of Ref.~\cite{goldstein}
for example). Then for a single eigenmode, the Schrodinger equation takes 
the form
\begin{equation}
\biggl [ 
- \left ( 1-\frac{R_S}{R} \right ) 
     \frac{1}{2m} \frac{\partial^2}{\partial b^2}
 + \frac{1}{2} K b^2 \biggr ] \psi (b,t)  = 
i \frac{\partial \psi (b,t)}{\partial t}
\label{schform}
\end{equation}
where $m$ and $K$ denote eigenvalues of ${\bf M}$ and ${\bf N}$,
and $b$ is the eigenmode. 

We re-write Eq.~(\ref{schform}) in the standard form 
\begin{equation}
\biggl [ 
- \frac{1}{2m} \frac{\partial^2}{\partial b^2}
 + \frac{m}{2} \omega^2 (\eta ) b^2 \biggr ] \psi (b,\eta )  
   = i \frac{\partial}{\partial \eta } \psi (b,\eta )
\label{shostandard}
\end{equation}
where
\begin{equation}
\eta  = \int^t_0 dt \left ( 1 - \frac{R_S}{R} \right )
\label{dtaudt}
\end{equation}
and
\begin{equation}
\omega^2 (\eta  ) = \frac{K}{m} \frac{1}{1-R_S/R}
                 \equiv \frac{\omega_0^2}{1-R_S/R}
\label{omegatau}
\end{equation}
We have chosen to set $\eta  (t=0)=0$. 

To proceed further, we need to choose the background spacetime
{\it i.e.} the behavior of $R(t)$. The classical solution
in Eq.~(\ref{solution}), tells us that $1-R_S/R \sim \exp(-t/R_S)$
at late times. We are mostly interested in the particle production
during this period. 
At early times, the behavior depends on how the
spherical domain wall was created and we are free to choose a
behavior for $R(t)$ that is convenient for calculations and
interpretation. To be able to interpret particle production
at very late times it is easiest to have a static situation. 
This can be obtained if we artificially take the collapse to
stop at some time, $t_f$. Eventually we can take $t_f
\rightarrow \infty$ to go over to the eternal collapse case.
So our choice for $R$ will be 
\begin{equation}
1- \frac{R_S}{R} = \left\{ \begin{array}{lll}
                    1  & & \mbox{$t \in (-\infty , 0)$} \\
                    e^{-t/R_S} & & \mbox{$t \in (0, t_f )$} \\
                    e^{-t_f /R_S}  & & \mbox{$t \in (t_f , \infty )$}
                            \end{array}
                    \right .
\end{equation}
This choice does have the issue that the derivative of $R$ has
discontinuities at $t=0$ and $t=t_f$. However, we shall show 
below that these discontinuities do not affect particle production.

\begin{figure}
\centerline{\scalebox{0.80}{\input{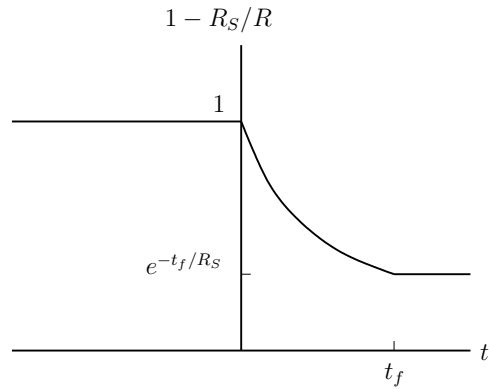}}}
\caption{Model for $R(t)$.
}
\label{modelRvst}
\end{figure}

With the chosen behavior of $R$, the spacetime is static at
early times and the initial vacuum state for the modes is the 
simple harmonic oscillator ground state, 
\begin{equation}
\psi (b,\eta =0) = \left ( \frac{m\omega_0}{\pi} \right )^{1/4}
              e^{-m\omega_0 b^2/2}
\label{psitau0}
\end{equation}
Then the exact solution to Eq.~(\ref{shostandard}) at later times is
\cite{Dantas:1990rk}
\begin{equation}
\psi(b,\eta) = e^{i\alpha (\eta)} 
            \left ( \frac{m}{\pi \rho^2} \right )^{1/4}
            \exp \left [ i \frac{m}{2} 
                   \left ( \frac{\rho_\eta}{\rho} +
                           \frac{i}{\rho^2} \right ) b^2
                 \right ]
\label{psisolution}
\end{equation}
where $\rho_\eta$ denotes derivative of $\rho (\eta)$ with 
respect to $\eta$, and $\rho$ is given by the real solution 
of the ordinary (though non-linear) differential equation
\begin{equation}
{\rho_{\eta\eta}} + \omega^2 (\eta) \rho = \frac{1}{\rho^3}
\label{rhoeq}
\end{equation}
with initial conditions
\begin{equation}
\rho (0) = \frac{1}{\sqrt{\omega_0}}\ , \ \ \ 
{\rho_\eta}(0) =0
\end{equation}
The phase $\alpha$ is given by
\begin{equation}
\alpha (\eta ) = - \frac{1}{2} \int_0^\eta 
                  \frac{d\eta'}{\rho^2 (\eta')}
\end{equation}
In Appendix~\ref{rhoeqdiscussion} we discuss the behavior of
$\rho$ as $\eta \rightarrow R_S$ ($t \rightarrow \infty$).
Also note that the solution for $\rho$ and $\rho_\eta$
is continuous.

Consider an observer with detectors that are designed to 
register particles of different frequencies 
for the free field 
$\phi$ at early times. Such an observer will interpret the 
wavefunction of a given mode $b$ at late times in terms of 
simple harmonic oscillator states, \{$\varphi_n$\}, at the 
{\em final} frequency, 
\begin{equation}
{\bar \omega} = \omega_0 e^{t_f /2R_S} 
\label{omegabar}
\end{equation}
The number of quanta in eigenmode $b$ can be evaluated by decomposing 
the wavefunction (Eq.~(\ref{psisolution})) in terms of the 
states, $\{ \varphi_n$\}, and by evaluating the occupation number 
of that mode. To implement this evaluation, we start by writing 
the wavefunction for a given mode at time $t > t_f$ in terms of the 
simple harmonic oscillator basis at $t=0$. 
\begin{equation}
\psi (b,t) = \sum_n c_n (t) \varphi_n (b) 
\end{equation}
where
\begin{equation}
c_n = \int db ~ \varphi_n^* (b) \psi (b,t)
\label{cndefn}
\end{equation}
which is an overlap of a Gaussian with the simple harmonic
oscillator basis functions. The occupation number 
at eigenfrequency ${\bar \omega}$ ({\it i.e.} in the eigenmode $b$) 
by the time $t > t_f$, is given by the expectation value
\begin{equation}
N (t , {\bar \omega}) = \sum_n n | c_n |^2 
\end{equation}

In Appendix~\ref{numberparticles} we evaluate the occupation
number in the eigenmode $b$ and the result is given in 
Eq.~(\ref{Nresult})
\begin{equation}
N (t , {\bar \omega}) = 
       \frac{{\bar \omega} \rho^2}{\sqrt{2}} \left [ 
      \left ( 1- \frac{1}{{\bar \omega} \rho^2} \right )^2
     + \left ( \frac{\rho_\eta}{{\bar \omega} \rho} \right )^2 
                \right ]
\end{equation}
for $t > t_f$.

By calculating ${\dot N}$, it can be checked that $N$ remains constant 
for $t < 0$ and also $t > t_f$. Hence all the particle production 
occurs for $0 \leq t \leq t_f$. There is a possibility that the particle
production is due to discontinuities in the derivative of $R$ at
$t=0, t_f$. However, as we shall see below, the particle number grows
with increasing $t_f$, while the discontinuity at $t=0$ is fixed, and
that at $t=t_f$ gets weaker. This indicates that 
particle production occurs only during $0 < t < t_f$ and is
a consequence of the gravitational collapse.

\begin{figure}
\scalebox{0.750}{\includegraphics{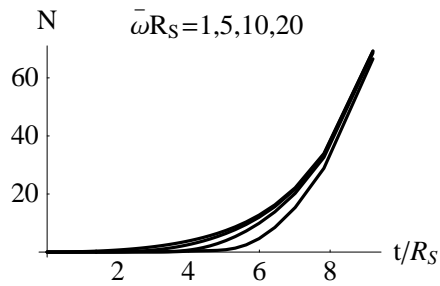}}
\caption{$N$ versus $t/R_S$ for various fixed values of 
${\bar \omega} R_S$. The curves are lower for higher
${\bar \omega}R_S$.
At late times the behavior is given by Eq.~(\ref{Ntanalytic}).
}
\label{Nvst}
\end{figure}

Now we can take the $t_f \rightarrow \infty$ limit. In 
Appendix~\ref{rhoeqdiscussion} we have shown that $\rho$ remains 
finite but $\rho_\eta \rightarrow - \infty$ as 
$t > t_f \rightarrow \infty$, provided $\omega_0 \ne 0$. 
However, we are interested in the behavior of $N$ for fixed 
frequency, ${\bar \omega}$. Since ${\bar \omega} = \omega_0 e^{+t_f/2R_S}$, 
$t_f \rightarrow \infty$ also implies $\omega_0 \rightarrow 0$. From the
discussion in Appendix~\ref{rhoeqdiscussion}, we also know that
$\rho \rightarrow \infty$ as $\omega_0 \rightarrow 0$. Hence we find 
\begin{equation}
N (t , {\bar \omega} ) \sim \frac{{\bar \omega} \rho^2}{\sqrt{2}}
\sim \frac{e^{t/(2 R_S)}}{\sqrt{2}} \ , \ \ \ t > t_f \rightarrow \infty
\label{Ntanalytic}
\end{equation}
This is confirmed by our numerical evaluation of $N$ as a
function of time $t > t_f$ for several different values of $\omega$
(see Fig.~\ref{Nvst}).

Therefore the occupation number at any frequency 
diverges in the infinite time limit when backreaction is not taken 
into account. This implies that backreaction due to particle creation
will have important consequences for gravitational collapse. 

\begin{figure}
\scalebox{0.7}{\includegraphics{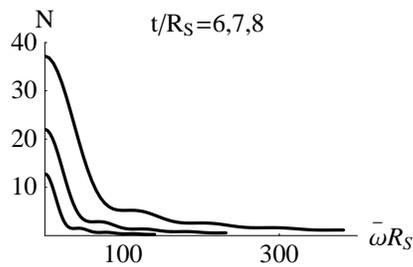}}
\caption{$N$ versus ${\bar \omega} R_S$ for various fixed values of 
$t/R_S$. The occupation number at any frequency grows as $t/R_S$ 
increases. 
}
\label{Nvsomega}
\end{figure}

\begin{figure}
\scalebox{0.7}{\includegraphics{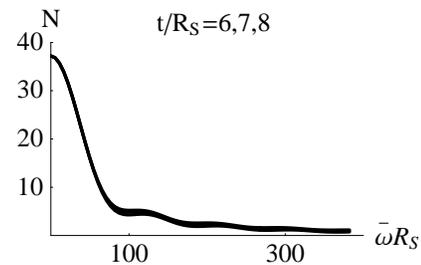}}
\caption{The same as Fig.~\ref{Nvsomega} but with the axes rescaled 
as in Eq.~(\ref{rescalinglaw}). This graph shows that the spectrum 
at different times is approximately self-similar.
}
\label{Nomegacollapsed}
\end{figure}

We have also numerically evaluated the spectrum of mode 
occupation numbers at any finite time and show the results in
Fig.~\ref{Nvsomega} for several different values of $t$. The
similar shapes of the different curves suggest that there
might be a simple relationship between them. By rescaling both 
axes we find that the curves roughly (though not completely)
collapse into a single
curve as shown in Fig.~\ref{Nomegacollapsed}. Hence, knowing
the spectrum at time $t_i$ approximately gives us the spectrum at all
times via
\begin{equation}
\lambda^{-1} (t) ~ N(t, {\bar \omega}/\lambda '(t)) = 
\lambda^{-1} (t_i) ~ N (t_i , {\bar \omega}/\lambda ' (t_i))
\label{rescalinglaw}
\end{equation}
where, we can determine the function $\lambda (t)$ by 
considering the time variation of $N(t, 0)$, and $\lambda '$ by
Eq.~(\ref{omegabar}). The result is
\begin{eqnarray}
\lambda (t) &=&  
\frac{1}{\sqrt{2}} \left [ e^{t/2R_S}+e^{-t/2R_S} -2 \right ] \\
\lambda ' (t) &=& e^{t/2R_S} 
\label{lambda}
\end{eqnarray}

We can compare the curve in Fig.~\ref{Nomegacollapsed} with the 
occupation numbers for the Planck distribution
\begin{equation}
N_P (\omega) = \frac{1}{e^{\beta \omega}-1}
\end{equation}
where $\beta$ is the inverse temperature. It is clear that 
the spectrum of occupation numbers is non-thermal. In particular, 
there is no singularity in $N$ at $\omega =0$ at finite time, 
there are oscillations 
in $N$, and the rescaling law of Eq.~(\ref{rescalinglaw}) is not 
applicable to a thermal distribution. As $t \rightarrow \infty$, 
the peak at $\omega =0$ does diverge and the distribution becomes more and more thermal. Even at finite
times, at small frequencies
\begin{equation}
N_P (\omega \ll \beta^{-1}) \approx \frac{1}{\beta \omega}
\label{smallfreqs}
\end{equation}
and the rescaling law amounts to rescaling the temperature by a 
factor $\lambda \lambda '$. 

\begin{figure}
\scalebox{0.750}{\includegraphics{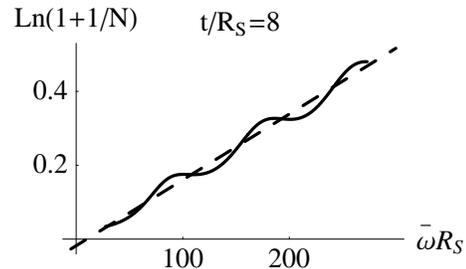}}
\caption{
${\rm Ln} (1+1/N)$ versus ${\bar \omega} R_S$ for $t = 8 R_S$. The 
dashed line shows ${\rm Ln} (1+1/N_P)$ versus ${\bar \omega} R_S$ 
where $N_P$ is a Planck distribution. The slope gives
$\beta^{-1}$ and the temperature in Eq.~(\ref{asympT}). 
}
\label{NandNP}
\end{figure}

Now, from Eq.~(\ref{shostandard}), since the time derivative of the
wavefunction on the right-hand side is with respect to $\eta$, $\omega$ 
is the mode frequency with respect to $\eta$ and not with respect to time 
$t$. Eq.~(\ref{dtaudt}) tells us that the frequency in $t$ is $(1-R_S/R)$ 
times the frequency in $\eta$, and at time $t_f$, this implies
\begin{equation}
\omega^{(t)} = e^{-t_f/R_s} {\bar \omega}
\end{equation}
where the superscript $(t)$ on $\omega$ refers to the fact that
this frequency is with respect to time $t$.
This rescaling of the frequency implies that the temperature for the
asymptotic observer (with time coordinate $t$) can be obtained by
find the ``best fit temperature'' $\beta^{-1}$ and then rescaling
by $(1-R_S/R)$. So the temperature seen by the asymptotic observer is 
\begin{equation}
T = e^{-t_f/R_S} \beta^{-1} (t_f)
\end{equation}
(The temperature $T$ is not to be confused with the time coordinate
within the spherical domain wall, also denoted by $T$ in Sec.~\ref{setup}.)
By using the scaling in Eq.~(\ref{lambda}), it is easy to see that
$\beta^{-1}$ grows as $e^{+t_f/2R_S}$ at late times and so $T$ is 
constant.  We can fit a thermal spectrum to the collapsed spectrum of
Fig.~\ref{Nomegacollapsed}, as shown in Fig.~\ref{NandNP} to obtain
\begin{equation}
T \approx \frac{0.19}{R_S} = \frac{2.4}{4\pi R_S} = 2.4 T_H
\label{asympT}
\end{equation}
where $T_H = 1/4\pi R_S \sim .08/R_S$ is the Hawking temperature. 
Since there is ambiguity in fitting the non-thermal spectrum by a 
thermal distribution, we can only say that the constant temperature, 
$T$, and the Hawking temperature are of comparable magnitude. 

The occupation number $N(t, \omega)$ can be related to the
asymptotic flux of radiation following standard procedures
(e.g. Chapter 8 of Ref.~\cite{BirrellandDavies}) and will
result in the usual greybody factors.

We thus see that in the context of the Schrodinger formalism 
there is evidence of Hawking-like, but non-thermal radiation 
emitted during gravitational collapse before any event horizon 
is formed. There are several possible sources that one can 
imagine for this radiation, including radiation due to a 
time-dependent metric, and also Hawking emission \cite{Hawking:1974sw}.  
Since the Schrodinger method in principle accounts for all such 
sources of radiation, it is worthwhile reexamining the original 
Hawking calculation, done using the Heisenberg picture and 
Bogolubov machinery, in the context of our above results.

\section{Hawking's calculation}
\label{appa}

In Hawking's pioneering paper \cite{Hawking:1974sw}, he considered 
a collapsing body. By matching asymptotic field operators, he could
find the Bogolubov coefficients, and then the particle emission
rate. The result is the famous Hawking thermal radiation at 
temperature
\begin{equation}
T_H = \frac{\kappa}{2\pi}
\end{equation}
where $\kappa = 1/2R_S$ is the surface gravity.

Since Hawking radiation is calculated in the $t \rightarrow
\infty$ limit (asymptotic field operators), the result does
not provide an answer to our original question: what will an
experimentalist observe at a finite time? So we must 
re-calculate the radiation from a collapsing domain wall which 
is close to, but still larger than, the Schwarzschild radius. 
Stated in a slightly different way -- does the experimentalist
see Hawking radiation before the event horizon is formed? 

\begin{figure}
\centerline{\scalebox{1.00}{\input{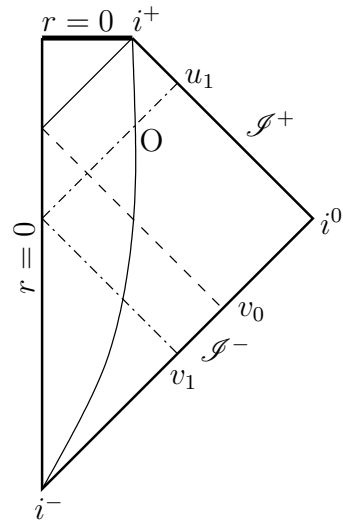}}}
\caption{Spacetime of a blackhole with null rays originating
at $\scri^-$ and going to $\scri^+$. The last ray that makes
it to $\scri^+$ is emitted at $v_0$. An observer, O, far from
the collapsing wall will attempt to detect a flux of radiation 
over a finite but large time interval.
The last ray to get to the observer originates at $\scri^-$
at $v_1 < v_0$ and arrives at $\scri^+$ at $u=u_1$. We are
interested in finding the particle flux in the section of
$\scri^+$ between the points marked $i^0$ and $u_1$.
}
\label{bh_spacetime}
\end{figure}

As Hawking showed, the mode functions of a massless scalar field 
in the black hole spacetime have a ``phase pile-up'' near the 
event horizon \cite{Hawking:1974sw}. In other words, 
if we retrace the mode functions from $\scri^+$ back in time 
up to $\scri^-$, the phase of the mode function diverges on 
$\scri^-$ at the point $v_0$ in Fig.~\ref{bh_spacetime}, 
where the coordinate $v$ is defined by
\begin{equation}
v = t+r + R_S \ln \left | \frac{r}{R_S} -1 \right | 
\end{equation}

The radial part of the ingoing mode functions on $\scri^-$ are
(Eq.~(2.11) of \cite{Hawking:1974sw})
\begin{equation}
f_{\omega '} = \frac{F_{\omega'}(r)}{\sqrt{2\pi\omega'}r} e^{i\omega' v}
\end{equation}
The relevant part of the outgoing mode function at frequency 
$\omega$ when extended back to $\scri^-$ is given in Eq.~(2.18) 
of \cite{Hawking:1974sw}
\begin{equation}
p_\omega^{(2)} \sim \frac{P_\omega^-}{\sqrt{2\pi \omega} r}
  \exp \left ( -i \frac{\omega}{\kappa} 
     \ln \left ( \frac{v_0-v}{CD} \right )
       \right ) \ , \ \ \ v < v_0
\label{modefn}
\end{equation}
and zero for $v > v_0$, where $P_\omega^-$, $C$, $D$ are 
constants, and $\kappa = 1/2R_S$. The expression in
Eq.~(\ref{modefn}) is only valid for small $v_0-v$,
and for large $\omega '$ (geometrical optics limit).

The overlaps of $p_\omega^{(2)}$ with $f_{\omega'}$ and 
${\bar f}_{\omega'}$ determine the Bogolubov coefficients.
This is equivalent to taking the Fourier transform of 
$p_\omega^{(2)}$. 
Following Hawking's calculation, 
the Bogolubov coefficients for large $\omega '$ are
(see Eq.~(2.19), (2.20) of \cite{Hawking:1974sw}; also see
\cite{Townsend})
\begin{eqnarray}
\alpha_{\omega\omega'}^{(2)} &\approx&
\frac{P_\omega^-}{2\pi} (CD)^{i\omega/\kappa} 
e^{i(\omega-\omega')v_0} \sqrt{\frac{\omega'}{\omega}}\nonumber \\
&& \times \Gamma \left ( 1- \frac{i\omega}{\kappa} \right ) 
(-i \omega')^{-1+i\omega/\kappa}  \\
\beta_{\omega\omega'}^{(2)} &\approx& 
-i \alpha_{\omega (-\omega')}^{(2)}
\end{eqnarray}

Even though the expression for $\alpha_{\omega\omega'}^{(2)}$
is only valid for large $\omega '$, Hawking argues on analyticity
grounds that the singularity at $\omega'=0$ should be present.
So to obtain $\beta_{\omega\omega'}^{(2)}$ it becomes necessary
to go around the pole at $\omega'=0$ to negative values of
$\omega'$. The choice of deformation of the contour around
the pole is determined on the grounds of analyticity, and
the result is
\begin{equation}
|\beta_{\omega\omega'}^{(2)}| =
|\alpha_{\omega(-\omega')}^{(2)}| =
\exp\left ( -\frac{\pi \omega}{\kappa} \right )
|\alpha_{\omega\omega'}^{(2)}| 
\label{Bogorelation}
\end{equation}
From here, the calculation of the thermal flux of Hawking
radiation follows.

\begin{figure}
\centerline{\scalebox{1.00}{\input{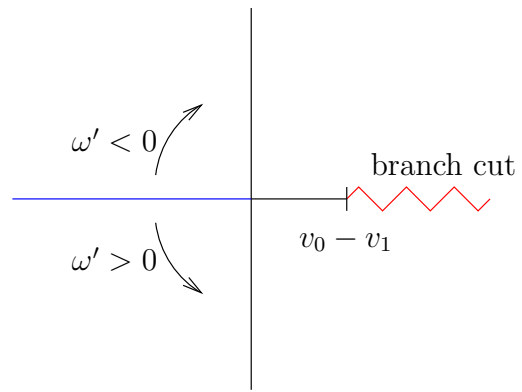}}}
\caption{The integration contour in the calculation of the 
Bogolubov coefficients runs from $v=-\infty$ to $v=0$ in the 
complex $v$ plane. In Hawking's calculation $v_0-v_1 =0$,
and the branch cut starts at the origin. For $\omega' <0$, the
integration contour is rotated to the upper imaginary axis,
and for $\omega' > 0$ to the negative imaginary axis. In
Hawking's case, this relates the Bogolubov coefficients
$\alpha_{\omega\omega'}$ and $\beta_{\omega\omega'}$ as
pedagogically described in Ref.~\cite{Townsend}.
(As pointed out to us by F. Dowker, care is required 
in comparing the calculations in \cite{Hawking:1974sw} and
\cite{Townsend} since Hawking considers modes 
$e^{+i\omega' v}$ while Townsend considers $e^{-i\omega' v}$.
We are following Hawking's calculation.) 
In our case, these very rotations can also be done. However, the branch
cut starts at $v_0-v_1 > 0$ and the simple relation between
the Bogolubov coefficients needed for thermal emission is not 
obtained.
}
\label{contours}
\end{figure}

Now consider an observer who only sees the collapsing object
for a finite time (see Fig.~\ref{bh_spacetime}). The last ray
detected by such an observer emerges from $\scri^-$ at 
$v=v_1 < v_0$. For this observer, the phase of the mode functions 
have a tendency to pile-up but there is no divergence as in 
Eq.~(\ref{modefn}) because $v \leq v_1 < v_0$. As far as this observer 
is concerned, the behavior in Eq.~(\ref{modefn}) holds for 
$v \leq v_1$, while for $v > v_1$ the back-tracked mode functions 
vanish. The Fourier transform of $p_\omega^{(2)}$ now gives the 
Bogolubov coefficients the following $\omega'$ dependence
\begin{equation}
\alpha_{\omega\omega'}^{(2)} \approx
\int^{v_1} dv \exp \left [ -i\omega' v -
              i\frac{\omega}{\kappa} \ln (v_0-v) \right ] 
\end{equation}
Following Ref.~\cite{Townsend}, for $\omega' > 0$ we rotate
the integration contour to the negative imaginary axis
($v \rightarrow -iy$) and for $\omega' < 0$ to the positive
imaginary axis (see Fig.~\ref{contours}). Simple manipulations 
for $\omega ' > 0$ then give
\[
|\alpha_{\omega\omega'}^{(2)}| =
e^{\pi\omega/2\kappa} \biggl | \int_0^\infty dy ~ 
       e^{-\omega' y - \omega \theta /\kappa}
         (y^2+\delta^2)^{i\omega/2\kappa}
\biggr |
\]
where $\delta =  (v_0-v_1)$ and $\theta = \tan^{-1}(\delta/y)$. 
Similarly, for $\omega' < 0$ we get
\[
|\alpha_{\omega\omega'}^{(2)}| \approx 
e^{-\pi\omega/2\kappa} 
\biggl | \int_0^\infty dy ~ 
       e^{-|\omega'| y + \omega \theta /\kappa}
         (y^2+\delta^2)^{i\omega/2\kappa}
\biggr |
\]
Since $\beta_{\omega\omega'}^{(2)} = \alpha_{\omega(-\omega')}^{(2)}$ 
the above expressions yield both Bogolubov coefficients.

The crucial difference between Hawking's asymptotic result and
the finite time result is the factor $\exp(\pm \omega\theta/\kappa)$
within the integral. Because of this factor, the relation in
Eq.~(\ref{Bogorelation}) does not hold and thermality is lost.
However, if this extra factor is nearly unity, we can expect the
spectrum to be nearly thermal. The integral is cut-off 
exponentially for $y > 1/\omega'$ and hence we estimate that
the spectrum will be nearly thermal provided 
$\omega \omega ' \delta /\kappa \ll 1$.
Hence the spectrum is thermal at low frequencies and gets closer
to being thermal as time goes on ($\delta \rightarrow 0$), both of 
which seem plausible on physical grounds.

It is difficult however, to go beyond these qualitative 
statements in attempting to compare our results with
what one might derive in the Hawking approach, in particular
to determine possibly how much of the effect we obtain might
be due to Hawking radiation, as opposed to particle creation
by a changing metric.  
This is because the spectrum depends on a sum over all $\omega '$, 
while the Hawking analysis is done in the geometrical optics limit, 
at large frequencies. Hence to find the spectrum in this approach, 
we need a more complete solution to the equations of motion 
for all the modes of the scalar field in the domain wall background. 
Such solutions are more difficult to obtain (as described in 
\cite{BirrellandDavies} for example).

\section{Infalling Observer}
\label{infallingobserver}

So far we have considered the wall collapse from the point of
view of an asymptotic observer. From the point of view of an
infalling observer, the time coordinate is $\tau$ of 
Sec.~\ref{setup} and the collapse appears to proceed differently. 
For example, if we ignore radiation, the classical equation of 
motion can be
written from the conservation of $M$ in Eq.~(\ref{MRtau}). 
Then, as the wall approaches the Schwarzschild radius,
\begin{equation}
R_\tau^2 \approx 
 \left [ \frac{M}{4\pi \sigma R_S^2} + 2\pi G\sigma R_S \right ]^2 -1
\end{equation}
The right-hand side is a non-zero constant, implying that the wall
is collapsing with constant velocity in the $\tau$ coordinate.
This shows that the collapse into a black hole occurs 
in a finite time interval for the infalling observer.
Further, Hawking has argued \cite{Hawking:1974sw}
that the infalling observer does
not detect significant Hawking radiation since the emission is
dominantly at low frequencies compared to $1/R_S$, while the
infalling observer can only have local detectors of size less
than $R_S$. Thus the infalling observer would appear to see 
event horizon formation in a finite time, with no significant 
radiation emanating from the black hole.

\begin{figure}
\centerline{\scalebox{0.80}{\input{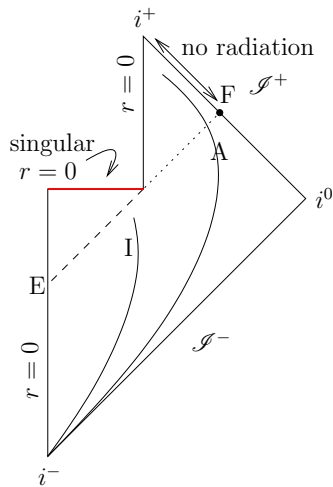}}}
\caption{The conventional spacetime diagram for an evaporating
black hole. 
The observer A will register a flux of quantum radiation even 
during collapse, and will be able to account for the entire energy 
of the shell by the time he/she gets to the line EF. From this point 
on, A will conclude that there is no energy left in the region of 
the collapsing shell. Yet A will see objects and other observers 
(such as I) disappear into what can at most be a Planck scale object 
and this is a puzzling feature of this picture.
}
\label{usualbh}
\end{figure}

These paradoxical views of the asymptotic and infalling observers
need to be reconciled, and the conventional way to reconcile
them is summarized in the spacetime diagram of an evaporating 
black hole shown in Fig.~\ref{usualbh}. The diagram is drawn so 
that the asymptotic observer sees evaporation in a finite time 
and the infalling observer falls into the black hole in a finite 
time also.

We have to note that the diagram in Fig.~\ref{usualbh} does not
follow from a rigorous solution to the problem of radiation from a
collapsing object with backreaction included. There are some
analyses of this problem in $(1+1)$-dimensional models
\cite{Davies:1976ei,Callan:1992rs} whose connection with the
$(3+1)$-dimensional problem is unclear (e.g.  \cite{dra}). 
Thus, the diagram in Fig.~\ref{usualbh} is a conjectured diagram 
that is widely used in the literature. While this diagram may 
well be the correct one once the full problem of gravitational 
collapse with backreaction is solved, we have to emphasize that 
it also has some puzzling features that indicate that it may not 
be the best conjecture to make in the absence of a backreaction 
analysis.

The conventionally drawn spacetime of an evaporating black hole 
has features that are not consistent with our findings. 
Since the asymptotic observer sees Hawking-like radiation 
from the collapsing wall {\it prior} to event horizon formation, the
mass of the collapsing wall must be decreasing, and at the point
denoted by $F$ in Fig.~\ref{usualbh} the entire energy of the wall 
has been radiated to $\scri^+$. However, in the spacetime of 
Fig.~\ref{usualbh}, it is at precisely this instant that the 
asymptotic observer sees infalling objects disappear into the
event horizon, even though there is nothing left of the collapsing 
wall to form the singularity.
A spacetime region such as the triangular region behind the event
horizon only seems reasonable if not all of the collapsing shell
energy has been lost to $\scri^+$ up to the point F, and there is 
some energy-momentum source left behind to crunch up in the singularity. 
Also, if the spacetime near the event horizon is described
by the Schwarzschild metric, there is infinite gravitational
redshfit of signals escaping to infinity, while the diagram
shows that signals escape to infinity in a finite time.
Finally, as is well known, the diagram in Fig.~\ref{usualbh}
also gives rise to the information loss paradox.
While these features of the diagram in Fig.~\ref{usualbh} are 
not inconceivable, they are sufficiently strange as to cast
doubt on the validity of the picture. 

Instead it may happen that the true event horizon never forms in a
gravitational collapse.  We saw that an outside observer never sees
formation of a horizon in finite time, not even in the full quantum
treatment. What about an infalling observer?  As in Hawking's case, 
the infalling observer 
does not see radiation, but this is due to size limitations of
his detectors. The mode occupation numbers we have calculated
will also be the mode occupation numbers that the infalling
observer will calculate, even if they be associated with frequency 
modes that he cannot personally detect. The infalling observer 
never crosses an event horizon, not because it takes an infinite 
time, but because there is no event horizon to cross. As the infalling
observer gets closer to the collapsing wall, the wall shrinks due
to radiation back-reaction, evaporating before an event horizon
can form. The evaporation appears mysterious to the infalling
observer since his detectors don't register any emission from
the collapsing wall. Yet he reconciles the absence of radiation
with the evaporation as being due to a limitation of the frequency
range of his detectors. Both he and the asymptotic observer would 
then agree that the spacetime diagram for an evaporating black hole
is as shown in Fig.~\ref{minkowski}. In this picture a global event
horizon and singularity never form. A trapped surface (from within
which light cannot escape) may exist temporarily, but after all of 
the mass is radiated, the trapped surface disappears and light gets 
released to infinity.

The spacetime picture that we are advocating is similar to that 
described in Refs.~\cite{Boulware:1975fe,Gerlach:1976ji} and, 
more recently, Refs.~\cite{Hajicek:1986hn,Ashtekar:2005cj,Alberghi:2001cm}.

\section{Discussion}
\label{discussion}

In this paper we have studied the collapse of a gravitating 
spherical domain wall using the functional Schrodinger equation. 
We would like to clearly delineate our analysis of the collapse and 
the emitted non-thermal quantum radiation from the interpretational
issues about the formation of an event horizon. 

First, we studied the collapse of a gravitating spherical domain
in both classical and quantum theory, ignoring any evaporative
processes. It has been suggested in the literature that quantum 
fluctuations can cause the collapse and formation of a black
hole in a finite (Schwarzschild) time \cite{frolov}. However,
our results show that this is not the case and the horizon does 
not form in a finite time even in the full quantum treatment. 

Then we studied radiation from the collapsing shell as seen by the 
asymptotic observer. In the process of gravitational collapse, there 
are two, perhaps related, sources of radiation: first is the radiation 
from particle creation in the changing gravitational field of the 
collapsing ball, and second may 
be Hawking-like  radiation due to a mismatch of vacua at early and late 
times. The functional Schrodinger analysis takes all such sources into
account and therefore gives the total particle production. We have found
a non-thermal distribution of particle occupation numbers,
with departures from thermality as illustrated in Fig.~\ref{Nvsomega}
and discussed toward the end of Sec.~\ref{semiclassradn}.
In a limited range of frequencies, the spectrum is approximately
thermal and the temperature fitted in a restricted range of
frequencies is constant and roughly equal to the Hawking temperature 
$1/4\pi R_S$. The radiation becomes thermal in the entire range of 
frequencies only in the limit $t \rightarrow \infty$, {\it i.e.}
when the horizon is formed.
Further, the mode occupation number diverges in the infinite time limit, 
if the backreaction is neglected ({\it i.e.} the background is held fixed).
Since an outside  observer never sees formation of a horizon in a finite 
time, radiation observed by him is never quite thermal. (Non-thermal 
features also get greatly amplified once the background is also treated 
quantum mechanically \cite{Vachaspati:2007hr}.)
This non-thermal radiation has strong implications for the information 
loss paradox since it can carry information about the collapsing matter.  

Without a rigorous calculation that includes backreaction, 
one can not give a definite answer to the final fate of a 
collapsing object. It may happen that the diagram in Fig.~\ref{usualbh} 
is correct and some radical and elaborate solutions to the problems we 
mentioned in Sec.~\ref{infallingobserver} are needed. However, 
one can imagine an alternative picture, different from the one in 
Fig.~\ref{usualbh}, which seems to have fewer problems, and that is 
that an event horizon never forms. Since
the mass of the shell is decreasing during the collapse, the shell
will be chasing its own Schwarzschild radius, and the question is
whether the shell will catch up to its own Schwarzschild radius or 
completely evaporate before that happens \cite{Gerlach:1976ji}.

\begin{figure}
\centerline{\scalebox{0.80}{\input{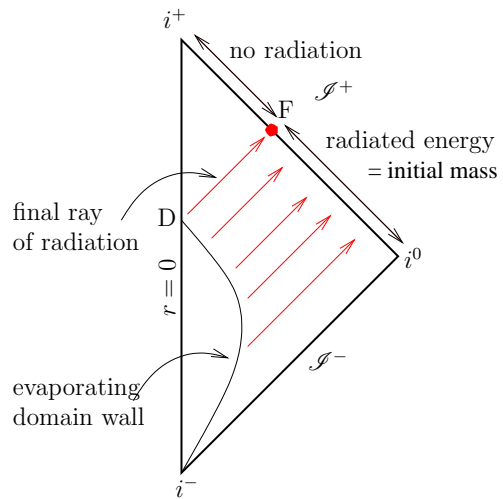}}}
\caption{The spacetime of a collapsing domain wall. During collapse
the wall emits non-thermal (quasi-Hawking) radiation as depicted by
the arrows. Our calculations indicate that the total energy
flux between the point $i^0$ to some point indicated by F is equal
to the energy of the initial domain wall. Hence we conjecture that
the domain wall evaporates completely at point D. Between F and $i^+$,
there is no radiation flux arriving at $\scri^+$. The event horizon
and singularity present in the customary treatment are not formed
and the spacetime structure is the same as that of Minkowski spacetime.
}
\label{minkowski}
\end{figure}

With backreaction included, the radiation should lead to a continual 
reduction of the Schwarzschild radius, $R_S$, occurring in the Ipser-Sikivie 
metric (see Sec.~\ref{setup}). Then, as seen by the asymptotic observer, 
one of two possibilities occurs: either the collapsing domain wall 
evaporates and $R_S \rightarrow 0$ in a finite time, 
or else backreaction causes the radiation rate to slow down and vanish
in a finite time. This latter possibility is unlikely, as our estimates 
suggest that the rate of emission increases as $R_S$ decreases \footnote{
Though our analysis only holds for objects of mass greater than Planck mass
(see Sec.~\ref{qcollapse}) and there could be qualitative changes in
the collapse and radiation as the Planck mass is approached.}.
We therefore conjecture that the backreaction due to particle production
will cause the collapsing domain wall spacetime to completely evaporate 
in a finite time. In this case, the spacetime can either be as 
given in Fig.~\ref{usualbh}, or have the same global spacetime structure 
as Minkowski space, as shown in Fig.~\ref{minkowski}. If the latter
picture is correct, it also means that the infalling observer will not 
encounter an event horizon, because this feature is simply absent from 
the spacetime.  
Another way to see this is to note that the causal relation between two
events is the same for all observers. Hence if the asymptotic observer 
sees a signal from an infalling object after he sees the last radiation 
ray emitted by the evaporating wall, this will also be the sequence of 
signals seen by the infalling observer. As discussed in 
Sec.~\ref{infallingobserver}, the infalling observer
would expect to see an intense burst of radiation as the wall approaches
the Schwarzchild radius, but can fail to do so because his detectors
are too small to detect the emitted range of frequencies.

In the absence of an exact backreaction calculation, we also have to 
allow for the possibility that a value of the critical mass exists
above which Fig.~\ref{usualbh} applies and below which Fig.~\ref{minkowski}
holds. Also, as discussed by Hawking \cite{Hawking:2005kf}, the question
of ``whether a black hole forms'' is not sharp enough and
may not make sense in the full quantum theory since all of the 
measurements are made by an asymptotic observer at infinity, while 
a collapsing object exists for a finite time and disappears by emitting 
radiation in the strong field region in the middle. An asymptotic observer 
can never be sure if a black hole formed because of underlying quantum 
uncertainty \cite{Hawking:2005kf}. 

The broad picture we have obtained is consistent with that proposed
in Refs.~\cite{Gerlach:1976ji,Boulware:1975fe}, though there are 
differences in the analysis and the conclusions. In particular, we 
find a non-thermal spectrum whereas Gerlach argues for thermality. 
Our picture also 
supports the interpretation of Hawking radiation given in
Ref.~\cite{Townsend} whereby particles
are created during the process of gravitational collapse and are
then radiated slowly to form what we call
Hawking radiation. We have indeed found particle production during
the collapse but the radiation is not quite thermal. 
It is only in the frequency range where the occupation number spectrum 
can be approximated by $T/\omega$ (Eq.~(\ref{smallfreqs})) that 
thermality holds at finite time. Also note that the non-thermality 
we find is in the mode occupation numbers. Propagation of the radiation 
in the background metric will cause further non-thermality due to 
greybody factors.

If we live in a world of low scale gravity, the collision of particles 
in high energy accelerators will lead to a situation where the particles 
are in a continual state of gravitational collapse from which 
non-thermal radiation is being emitted. The life-time of such a state 
can be estimated once we know the details of the radiation more precisely
from an analysis which includes backreaction. 
However, on dimensional grounds, Hawking's estimate for the lifetime of
a black hole ($\sim R_S^3/G$) may well apply to the colliding
particles as well.

In reality the collapse is further complicated by the fact that the 
collapsing object is not kept in isolation and there are external forces 
that can disrupt the collapse at any point in time. 
From the perspective of potential information loss, note that
any infalling encyclopedias 
can be returned to the asymptotic observer if the collapse
is disrupted at any time, as it could be, for example by
a bomb set to go off at some late, but finite time. 
Most importantly, since we calculate that 
the radiation emitted during the
gravitational collapse is never truly thermal, the classic information 
loss issue in black holes should, in this case, be a non-problem for 
the asymptotic observer \footnote{To be sure, one must explicitly 
evaluate the information contained in the non-thermal radiation.}.

Our primary result, that no event horizon forms in gravitational
collapse as seen by an asymptotic observer is suggestive of the
possibility of using the number of local event horizons to classify 
and divide Hilbert space into superselection sectors, labeled by the 
number of local event horizons.  Our result suggests that no operator 
could increase the number of event horizons, but the possibility of 
reducing the number of pre-existing primordial event horizons is not 
so clear and would require that Hawking radiation not cause any 
primordial black hole event horizons to evaporate completely.  

Our conclusions have been derived on the basis of a number of assumptions 
which we now discuss. The first is the truncation of superspace to 
minisuperspace. We have only included spherically symmetric field 
configurations. Even then, the metric is restricted to be the classical 
solution sourced by a spherical domain wall. A more general analysis would 
include more metric degrees of freedom, though it is hard to see how
this would make a difference to our conclusion. Similarly, we have 
restricted ourselves to a zero thickness domain wall. A more general 
analysis would allow for a thick wall. Finally, the Wheeler-de Witt 
formalism as we have used it, does
not allow for the creation and annihilation of domain walls (``third
quantization''). Perhaps third quantization could allow for the 
spontaneous creation of a black hole and the annihilation of the
wall, effectively leading to black hole formation. 
A fourth possibility is that our Lagrangian breaks down near the 
Schwarzschild horizon and ``quantum gravity'' effects become important. 
This is usually thought not  to be the case since the spacetime curvature 
near the horizon is small for large black holes. 

Perhaps the most serious drawback of our analysis is that it
does not include backreaction on the gravitational collapse due 
to radiation. While we do not expect such inclusion to alter
our conclusions regarding the non-existence of event horizons for 
asymptotic observers, we  are currently exploring ways to 
extend our treatments to include backreaction.

No theoretical idea is complete without the possibility of 
experimental verification and so it is important to ask if
the picture we have developed in this paper can also be
tested experimentally. We have already mentioned the relevance
of our conclusions to black hole production in particle
accelerators provided low scale gravity is correct. However,
there is an even more accessible experimental system where
these theoretical ideas can be put to the test. These are
condensed matter systems in which sonic black holes (dumbholes)
may exist \cite{Unruh:1980cg}.
It is very hard to realize a dumbhole in the laboratory 
for various experimental reasons; the closest known realization
seems to be the propagating He-3 AB interface in the experiment 
of Ref.~\cite{Baretal00} as discussed in \cite{Vac04}.
Yet the crucial aspect of our work in this paper is
that {\em there is no need to produce a dumbhole in order 
to see acoustic ``pre-Hawking'' radiation}. The process of
collapse toward a dumbhole will give off radiation. This is
also the conclusion of Ref.~\cite{Barceloetal} though the
details of the analysis and conclusions are different -- for
example, we find non-thermal emission whereas these authors
claim thermal emission with a modified temperature that is lower
than the Hawking temperature. In any case, it should be much easier 
to do experiments in the laboratory that do not go all the way to 
forming a dumbhole, and this could be an ideal arena to test 
pre-Hawking radiation.

Our conclusions are important not only for the general issue of the 
breakdown of unitarity via information loss, but also for more general 
studies of black hole formation, whether they be in the context of 
astrophysics ({\it e.g.} galactic black holes) or in future accelerator 
experiments. 
In all these situations, we are asymptotic observers watching the 
gravitational collapse of matter, and we may never see effects associated 
with a black hole event horizon. Only effects occurring during the 
gravitational collapse itself appear to be visible.

\begin{acknowledgments}
We are grateful to Fay Dowker, Larry Ford, Alan Guth, Jonathan Halliwell, 
Irit Maor, Harsh Mathur, Don Page, Paul Townsend, Alex Vilenkin, Bob Wald,
Frank Wilczek, and Serge Winitzki for their interest, advice, and feedback. 
TV is also grateful to the participants of the COSLAB meeting at the Lorentz 
Center (Leiden), including Brandon Carter, Tom Kibble, Bill Unruh and 
Matt Visser, for discussion. TV acknowledges 
hospitality by Imperial College and Leiden University where some of this 
work was done.  DS is grateful to participants of COSMO 06 meeting for very useful comments and discussions. 
LMK acknowledges hospitality of
Vanderbilt University during the completion of this work.
This work was supported by the U.S. Department of Energy, 
NASA at Case Western Reserve University, and NWO (Netherlands) at Leiden
University. 
\end{acknowledgments}

\appendix

\section{$\rho$ equation}
\label{rhoeqdiscussion}

In the range $t < 0$, $\omega$ is a constant and the solution
to Eq.~(\ref{rhoeq}) is
\begin{equation}
\rho (\eta ) = \frac{1}{\sqrt{\omega_0}}
\end{equation}

In the range $0< t < t_f$, we do not have an analytic solution
but we can derive certain useful properties. First note that
in terms of $\eta$
\begin{equation}
\omega^2 = \frac{\omega_0^2}{1-\eta/R_S}
\end{equation}
Then the equation for $\rho$ after rescalings can be written as:
\begin{equation}
\frac{d^2 f}{d{\eta '} ^2} = 
- (\omega_0 R_S )^2 \left [ \frac{f}{1-{\eta '}} - \frac{1}{f^3} \right ]
\label{feq}
\end{equation}
where ${\eta '} = \eta /R_S$, $f = \sqrt{\omega_0} \rho$.
The boundary conditions are
\begin{equation}
f(0) = 1 \ , \ \ \ \frac{df(0)}{d{\eta '}} = 0
\end{equation}
The last term with the $1/f^3$ becomes singular as $f \rightarrow 0$.
Let us consider another equation with the $1/f^3$ replaced by something
better behaved. For example,
\begin{equation}
\frac{d^2 g}{d{\eta '}^2} = 
- (\omega_0 R_S )^2 \left [ \frac{g}{1-{\eta '}} - g \right ]
\label{softereq}
\end{equation}
with boundary conditions 
\begin{equation}
g(0) = 1 \ , \ \ \ \frac{dg(0)}{d{\eta '}} = 0
\label{softbcs}
\end{equation}
Eq.~(\ref{softereq}) implies that $g(\eta ')$ is monotonically
decreasing as long as $g(\eta ') > 0$. Furthermore, it is
decreasing faster than the solution for $f$ as long as $f < 1$,
since the $1/f^3$ term in Eq.~(\ref{feq}) is a larger ``repulsive'' 
force than the $g$ term in the Eq.~(\ref{softereq}). So
\begin{equation}
g({\eta '} ) \leq f({\eta '} )
\end{equation}
for all $\eta '$ such that $g(\eta ' ) > 0$. 

Eqs.~(\ref{softereq}) with initial conditions (\ref{softbcs}) can be 
solved in terms
of degenerate hypergeometric functions. For us, the important point 
is that the solution for $g$ is positive for all ${\eta '}$ and,
in particular, $g (1) > 0$ for all the values of $\omega_0 R_S$
that we have checked. 
Therefore $f({\eta '})$ is positive, at least for a wide range of 
$\omega_0 R_S$.
        
Let $f_1 = f(1) \ne 0$. Then the equation for $f$ can be expanded
near ${\eta '}=1$.
\begin{equation}
\frac{d^2f}{d{\eta ' }^2} \sim - (\omega_0 R_S)^2 
                  \left [ \frac{f_1}{1-{\eta '}} - \frac{1}{f_1^3} \right ]
\end{equation}
This shows that 
\begin{equation}
\frac{df}{d{\eta '}} \sim  (\omega_0 R_S)^2 f_1 
                          \ln (1-{\eta '}) \rightarrow - \infty
\end{equation}
as ${\eta '} \rightarrow 1$.

Hence $\rho(\eta = R_S)$ is strictly positive and finite while
$\rho_\eta (\eta = R_S) = -\infty$ for finite and non-zero
$\omega_0$. Since $f = \sqrt{\omega_0} \rho$, and $f \rightarrow
1$ for $\omega_0 \rightarrow 0$, we also see that 
$\rho \rightarrow \infty$ and $\rho_\eta \rightarrow 0$
as $\omega_0 \rightarrow 0$.

In the range $t_f < t$, $\omega$ is a constant. However, the
solution for $\rho$ is not a constant, unlike in the range
$t < 0$, since the constant solution $1/\sqrt{\omega (t_f)}$
does not necessarily match up with $\rho (t_f -)$ to ensure
a continuous solution. Yet it is easy to check that in this
region ${\dot N} = 0$ and so there is no change in the
occupation numbers. So we need only find $N(t_f-, {\bar \omega})$
to determine $N(t \rightarrow \infty , {\bar \omega})$.

\section{Number of particles radiated as a function of time}
\label{numberparticles}

We use the simple harmonic oscillator basis states but at a
frequency ${\bar \omega}$ to keep track of the different
$\omega$'s in the calculation. To evaluate the occupation
numbers at time $t > t_f$, we need only set 
${\bar \omega}= \omega (t_f)$. So
\begin{equation}
\phi_n (b) = \left ( \frac{m{\bar \omega}}{\pi} \right )^{1/4}
        \frac{e^{-m{\bar \omega} b^2/2}}{\sqrt{2^n n!}}
          H_n (\sqrt{m{\bar \omega}} b)
\end{equation}
where $H_n$ are Hermite polynomials.
Then Eq.~(\ref{cndefn}) together with Eq.~(\ref{psisolution}) gives
\begin{eqnarray}
c_n &=& \left ( \frac{1}{\pi^2 {\bar \omega} \rho^2} \right )^{1/4}
 \frac{e^{i\alpha}}{\sqrt{2^n n!}} 
 \int d\xi e^{-P\xi^2 /2} H_n (\xi ) \nonumber \\
 &\equiv& \left ( \frac{1}{\pi^2 {\bar \omega} \rho^2} \right )^{1/4}
 \frac{e^{i\alpha}}{\sqrt{2^n n!}} I_n
\label{cn}
\end{eqnarray}
where
\begin{equation}
P = 1 - \frac{i}{\bar \omega} 
       \left ( \frac{\rho_\eta}{\rho} + \frac{i}{\rho^2} \right )
\end{equation}

To find $I_n$ consider the corresponding integral over the
generating function for the Hermite polynomials
\begin{eqnarray}
J (z) &=& \int d\xi e^{-P\xi^2/2} e^{-z^2 + 2z\xi} \nonumber \\
        &=& \sqrt{\frac{2\pi}{P}} e^{-z^2 (1-2/P)}
\end{eqnarray}
Since 
\begin{equation}
e^{-z^2 + 2z\xi} = \sum_{n=0}^{\infty} \frac{z^n}{n!} H_n (\xi )
\end{equation}
\begin{equation}
\int d\xi e^{-P\xi^2 /2} H_n (\xi ) = 
          \frac{d^n}{dz^n} J (z) \biggr |_{z=0}
\end{equation}
Therefore
\begin{equation}
I_n = \sqrt{\frac{2\pi}{P}} \left ( 1- \frac{2}{P} \right )^{n/2}
                  H_n (0)
\end{equation}
Since
\begin{equation}
H_n (0) = (-1)^{n/2} \sqrt{2^n n!} \frac{(n-1)!!}{\sqrt{n!}} \ ,\ \ \ 
n = {\rm even}
\end{equation}
and $H_n (0)=0$ for odd $n$, we find the coefficients
$c_n$ for even values of $n$,
\begin{equation}
c_n = \frac{(-1)^{n/2} e^{i\alpha}}{({\bar \omega} \rho^2)^{1/4}} 
      \sqrt{\frac{2}{P}}  \left ( 1- \frac{2}{P} \right )^{n/2}
      \frac{(n-1)!!}{\sqrt{n!}}
\label{cnresult}
\end{equation}
{}For odd $n$, $c_n=0$.

Next we find the number of particles produced. Let
\begin{equation}
\chi = \biggl | 1- \frac{2}{P} \biggr |
\end{equation}
Then
\begin{eqnarray}
N (t ,{\bar \omega} ) &=& \sum_{n={\rm even}} n |c_n|^2 \nonumber \\
            &=& \frac{2}{\sqrt{{\bar \omega} \rho^2} |P|}
                 \chi \frac{d}{d\chi} 
	\sum_{n={\rm even}} \frac{(n-1)!!}{n!!} \chi^n \nonumber \\
            &=& \frac{2}{\sqrt{{\bar \omega} \rho^2} |P|}
                 \chi \frac{d}{d\chi} \frac{1}{\sqrt{1-\chi^2}} \nonumber \\
            &=& \frac{2}{\sqrt{{\bar \omega} \rho^2} |P|}
                  \frac{\chi^2}{(1-\chi^2)^{3/2}} 
\end{eqnarray}
Inserting the expressions for $\chi$ and $P$, leads to
\begin{equation}
N (t , {\bar \omega}) = 
       \frac{{\bar \omega} \rho^2}{\sqrt{2}} \left [ 
      \left ( 1- \frac{1}{{\bar \omega} \rho^2} \right )^2
     + \left ( \frac{\rho_\eta}{{\bar \omega} \rho} \right )^2 
                \right ]
\label{Nresult}
\end{equation}

In summary, we have found the occupation number of modes as a function
of $\rho$ which is a function of time as given by the non-linear 
differential equation Eq.~(\ref{rhoeq}). The equation connecting 
$\rho$ and time $t$ has only been solved numerically but we have 
discussed the behavior of $\rho$ and $\rho_\eta$ as 
$\eta \rightarrow R_S$ ($t \rightarrow \infty$) 
in Appendix~\ref{rhoeqdiscussion}.

\end{document}